\documentclass{article}

\usepackage{arxiv}

\usepackage[utf8]{inputenc} 
\usepackage[T1]{fontenc}    
\usepackage{hyperref}       
\usepackage{url}            
\usepackage{booktabs}       
\usepackage{amsfonts}       
\usepackage{nicefrac}       
\usepackage{microtype}      
\usepackage{cleveref}       
\usepackage{lipsum}         
\usepackage{graphicx}
\usepackage{natbib}
\usepackage{doi}
\usepackage{multirow}

\title{Statistical Post-Processing for Gridded Temperature Prediction Using Encoder–Decoder-Based Deep Convolutional Neural Networks}

\author{ \href{https://orcid.org/0000-0000-0000-0000}{\hspace{1mm}Atsushi Kudo} \\
	Numerical Prediction Development Center \\
	Japan Meteorological Agency \\
	Tsukuba, Japan \\
	\texttt{atsushi.kudou@met.kishou.go.jp } \\
}

\renewcommand{\shorttitle}{Gridded Temperature Prediction Using Deep Convolutional Neural Networks}

\hypersetup{
pdftitle={Statistical Post-Processing for Gridded Temperature Prediction Using Encoder–Decoder-Based Deep Convolutional Neural Networks},
pdfsubject={},
pdfauthor={Atsushi Kudo},
pdfkeywords={Deep convolutional neural networks, Statistical post-processing, Temperature forecast},
}

\begin{document}
\maketitle

\begin{abstract}
The Japan Meteorological Agency operates gridded temperature guidance to predict two-dimensional snowfall amounts and precipitation types, e.g., rain and snow, because surface temperature is one of the key elements to predict them. Operational temperature guidance is based on the Kalman filter, which uses temperature observation and numerical weather prediction (NWP) outputs only around observation sites. Correcting a temperature field when NWP models incorrectly predict a front’s location or when observed temperatures are extremely cold or hot has been challenging. 

In this study, an encoder–decoder-based convolutional neural network has been proposed to predict gridded temperatures at the surface around the Kanto region in Japan. Verification results showed that the proposed model greatly improves the operational guidance and can correct NWP model biases, such as a positional error of fronts and extreme temperatures. 
\end{abstract}


\section{Introduction}
\label{sec:sec1}
Numerical weather prediction (NWP) models have played a central role in issuing weather forecasts and information in recent decades \citep{WMO2013}. NWP centers across the world have been operating NWP models and working to improve prediction accuracy. Since NWP models have systematic errors \citep{Zadra2018, Hacker2007} because of their resolutions and limitations, statistical post-processing, or “guidance” \citep{Klein1974, Zurndorfer1979} has been applied to provide more accurate predictions by mitigating these systematic errors. Additionally, guidance is used to evaluate variables not directly derived by NWP models, e.g., probability of thunderstorms, weather category, and aviation turbulence \citep{JMA2018}. 

The Japan Meteorological Agency (JMA) has operated various guidance, for predicting temperature; wind; precipitation; snowfall amount; precipitation type, e.g., rain and snow; weather category; humidity; etc. \citep{JMA2018, JMA2019a}. Two types of temperature guidance have operated at the JMA: point-like temperature guidance and gridded temperature guidance. Point-like temperature guidance predicts sequential, maximum, and minimum temperatures at meteorological observatories in Japan on the basis of the Kalman filter technique \citep{Kalman1960}. On the other hand, gridded temperature guidance predicts sequential temperatures at each grid point on the basis of the point-like temperature guidance; that is, the gridded temperature guidance is calculated from the difference between point-like temperature guidance and NWP models at each observatory \citep{Furuichi2010}. The operational gridded temperature guidance has been used as input values for the snowfall amount and precipitation type guidance because surface temperature is one of the key elements to predict them. Thus, improving the gridded temperature guidance accuracy is crucial to improve predictions of snowfall amounts and precipitation types.

The Kalman filter is a sequential analysis method used in various industries, such as robotics, space technology, economy, and meteorology. For temperature guidance, the Kalman filter is used to estimate the coefficients of explanatory variables at the targeting time \citep{Persson1989, Simonsen1991}. Although the Kalman filter can greatly improve temperature predictions, the operational temperature guidance at the JMA cannot predict temperature accurately when NWP models, used as input variables, mispredict the position of fronts or predict extreme high or low temperatures \citep{Sannohe2018}. There are three main reasons for this difficulty. First, the current temperature guidance uses variables only around observatories so that the guidance cannot correct positional biases. Second, the Kalman filter is based on a linear Gaussian state-space model, so it cannot represent nonlinear relationships among predictors and predictand. Lastly, the current temperature guidance is adjusted to follow variations in several weeks; hence, it cannot follow the daily change of meteorological conditions.

Recent advancements in machine learning technologies, particularly deep learning (DL) or deep neural networks, have largely impacted various industries due to precise image and speech recognition capability \citep{Alam2020, Yu2020}. Many meteorological studies based on DL have been conducted in recent years, including the discrimination of typhoon intensity \citep{Pradhan2018, Wimmers2019, Daikoji2020}, precipitation nowcasting \citep{Shi2015, Samsi2019, Agrawal2019}, quality check algorisms for observed values \citep{Dai2018}, data assimilation \citep{Bocquet2019, Arcucci2021}, and subgrid parameterizations \citep{Rasp2018, Han2020}. However, only a few studies based on DL have been conducted on guidance for short-range weather forecasts \citep{Scheuerer2020, Veldkamp2021}, despite it having used machine learning techniques for approximately 50 years \citep{Glahn1972}.

In this study, deep encoder–decoder networks \citep{Badrinarayanan2017} are used to predict two-dimensional (2D) surface temperatures to 15-h forecast lead time around the Kanto region in Japan. The present study aims to show the proposed 2D temperature forecast surpasses the operational gridded temperature guidance, which would help improve the operational snowfall amount and precipitation type forecast. This paper is structured as follows. In section 2, the characteristics of input values used to train the network and to infer the temperature distribution map are described. Section 3 shows the neural network (NN) constructed in this study. Section 4 shows the results of verification using data independent from the training period. Section 5 presents the case studies, and section 6 mentions the conclusion and future issues.

\section{Input variables}
\label{sec:sec2}

\begin{figure}[b]
  \centering
  \includegraphics[width=0.76\textwidth,trim=0 190 0 0,clip]{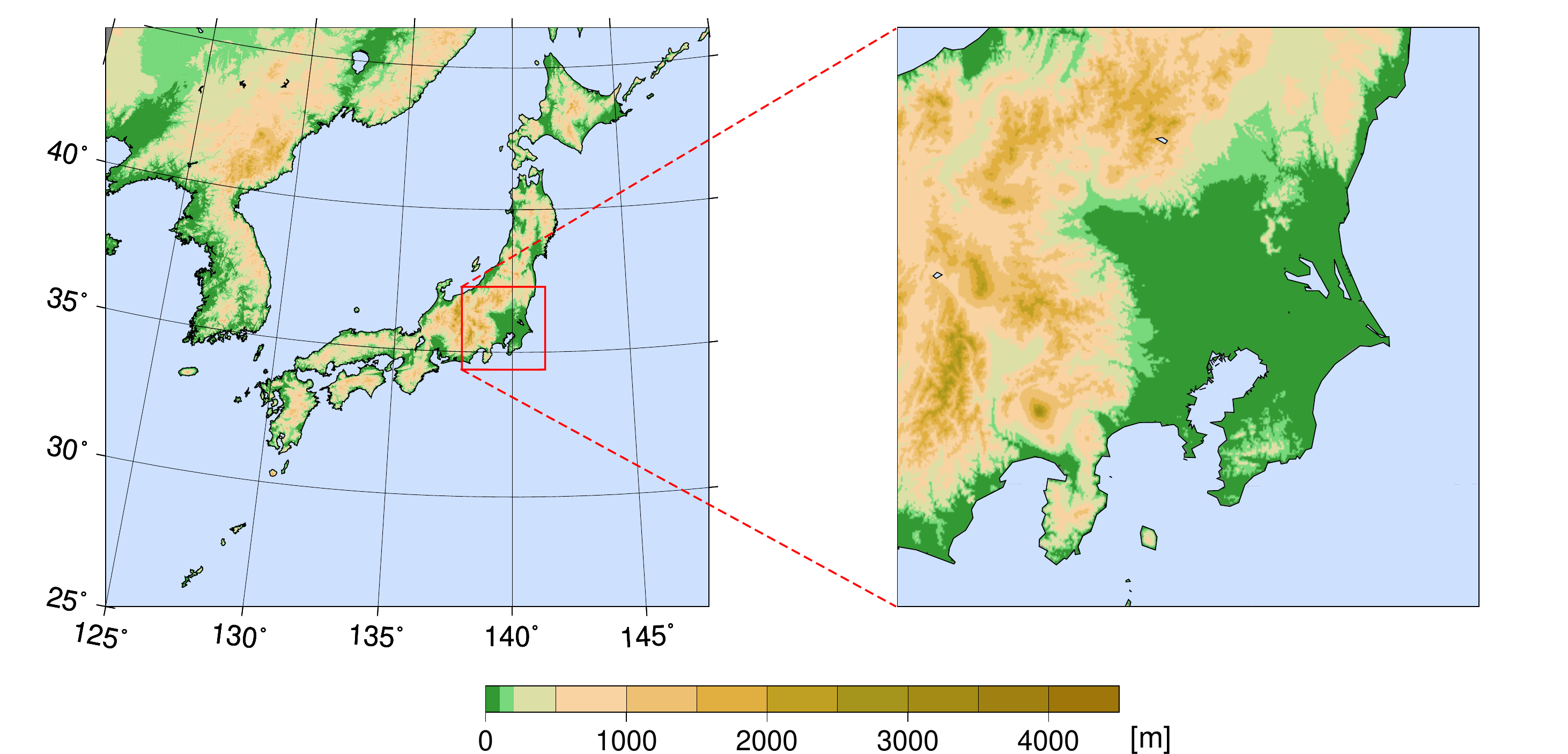}
  \caption{Japanese Islands (left) and the studied area (right). The color shades illustrate topography.}
  \label{fig:fig1}
\end{figure}

\subsection{NWP data}
\label{sec:sec2-1}

This study uses NWP model outputs from the JMA’s operational Meso-scale model (MSM, \cite{JMA2019b}), a non-hydrostatic regional model with 5-km grid spacing, as the input variables to the networks. The target area is set to $64\times 64$ grid points around the Kanto region to focus on a specific area and shorten the calculation time. The area is shown in the right panel in Fig. \ref{fig:fig1}. The input variables used in this study are seven MSM output variables, namely, temperatures at the surface; 975, 925, and 850 hPa; mean sea level pressure (MSLP); and surface wind components U and V, which are empirically selected using the training and validation dataset, described later. Forecast lead times used are 3, 6, 9, 12, and 15 h because the forecast length of MSM was 15 h as of 2010, which is the beginning year of the training dataset.

\subsection{Estimated temperature distribution as ground truth}
\label{sec:sec2-2}

The JMA has operated the estimated weather distribution products, providing real-time weather and surface temperature distribution with a 1-km grid spacing \citep{Wakayama2020}. In this study, temperature from the estimated weather distribution products is used as the ground truth of the network. The temperature distribution is analyzed using both the observed temperature, collected from approximately 930 Automated Meteorological Data Acquisition System (AMeDaS) sites around Japan, and the mesh-averaged temperature, a 30-year averaged temperature map from 1981 to 2010 that is estimated from latitude, longitude, height, gradient, and urban factors at each grid. Since February 2019, 1-h forecast temperatures from the Local forecast model (LFM, \cite{JMA2019b}), a rapid update model with 2-km grid spacings, have also been used in the analysis to improve the accuracy of the estimated temperature \citep{Wakayama2020}.

In this study, the estimated temperature was averaged in 5-km grids, along with MSM. Since the estimated temperature is calculated only on land, MSM surface temperature at the corresponding initial time is substituted on the sea to avoid unfavorable training results. As described in section 4, only the 2-m temperature over land will be used to verify prediction accuracy.

\section{Structure of the neural network}
\label{sec:sec3}

\begin{figure}[b]
  \centering
  \includegraphics[width=0.95\textwidth]{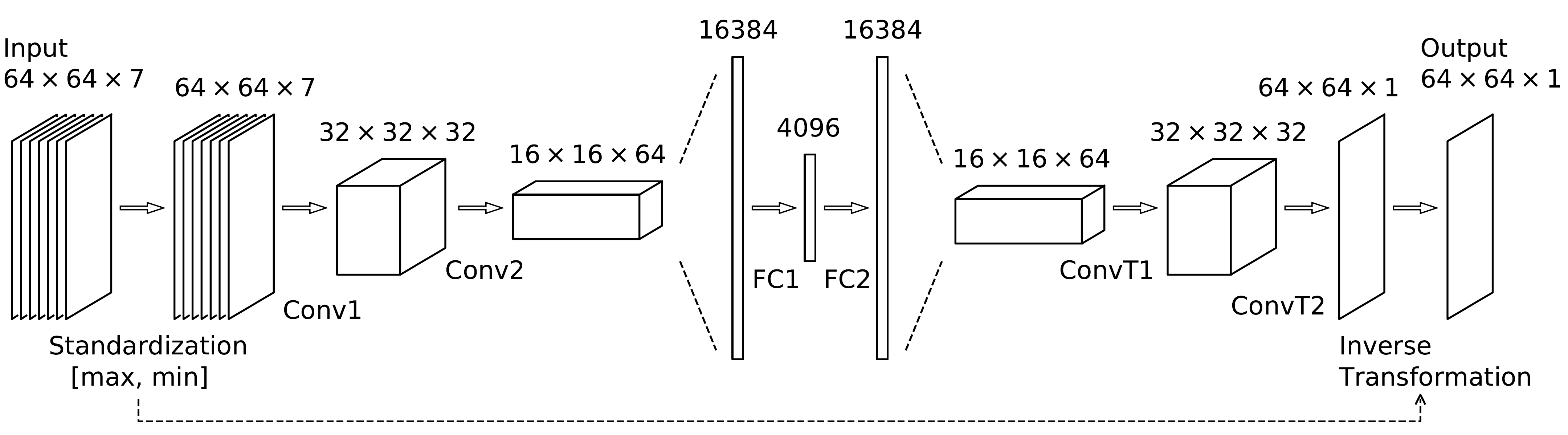}
  \caption{Schematic diagram of deep convolutional neural network used in this study. This network inputs seven channels of images with $64\times 64$ grid points and outputs an image with $64\times 64$ grid points. The maximum and minimum values of input images are used to standardize input images to a 0–1 value range and to inversely transform the output from the neural network model. The detail of operations, such as Conv1 and Conv2, in the figure is described in Table \ref{tab:tab1}.}
  \label{fig:fig2}
\end{figure}

\subsection{Encoder–decoder model}
\label{sec:sec3-1}

This study uses encoder–decoder-based NNs, hereafter referred to encoder–decoder models, to predict 2D temperatures. Encoder–decoder models input and output one or more channels of images with small-sized characteristic maps in the middle layer. Because encoder–decoder models’ construction allows networks to treat 2D variables, such as temperature, humidity, and wind, with some channels of NWP outputs as input values, these models are suitable for producing 2D guidance.

\begin{table}[h]
  \centering
  \caption{Functions and parameters used in the network shown in Fig. \ref{fig:fig2}}
  \vspace{-4mm}
  \label{tab:tab1}
  {\small
  \begin{tabular}[t]{|p{10mm}|p{25mm}|p{85mm}|}
    \hline
    Layer & Function & Parameters \\
    \hline
    \multirow{4}{*}{Conv1} & Conv2d & kernel\_size=5, stride=1, padding=2, number of channels: 7$\rightarrow$32 \\
    \cline{2-3}
    & MaxPool2d & kernel\_size=2, stride=2 \\
    \cline{2-3}
    & BatchNorm2d & number of channels: 32 \\
    \cline{2-3}
    & ReLU & \\
    \hline

    \multirow{4}{*}{Conv2} & Conv2d & kernel\_size=5, stride=1, padding=2, number of channels: 32$\rightarrow$64 \\
    \cline{2-3}
    & MaxPool2d & kernel\_size=2, stride=2 \\
    \cline{2-3}
    & BatchNorm2d & number of channels: 64 \\
    \cline{2-3}
    & ReLU & \\
    \hline

    \multirow{3}{*}{FC1} & Linear & number of units: 16384$\rightarrow$4096 \\
    \cline{2-3}
    & BatchNorm1d & number of channels: 4096 \\
    \cline{2-3}
    & ReLU & \\
    \hline

    \multirow{3}{*}{FC2} & Linear & number of units: 4096$\rightarrow$16384 \\
    \cline{2-3}
    & BatchNorm1d & number of channels: 16384 \\
    \cline{2-3}
    & ReLU & \\
    \hline

    \multirow{3}{*}{ConvT1} & ConvTranspose2d & kernel\_size=2, stride=2, padding=0, number of channels: 64$\rightarrow$32 \\
    \cline{2-3}
    & BatchNorm2d & number of channels: 32 \\
    \cline{2-3}
    & ReLU & \\
    \hline

    \multirow{3}{*}{ConvT2} & ConvTranspose2d & kernel\_size=2, stride=2, padding=0, number of channels: 32$\rightarrow$1 \\
    \cline{2-3}
    & BatchNorm2d & number of channels: 1 \\
    \cline{2-3}
    & Sigmoid & \\
    \hline

  \end{tabular}
  }
\end{table}

\subsection{Neural network used in the study}
\label{sec:sec3-2}

The network used in this study is a deep encoder–decoder-based convolutional neural network (CNN), as shown in Fig. \ref{fig:fig2}. The network inputs seven channels of NWP outputs, and then, they are standardized into a 0–1 value range using the maximum and minimum value, discussed in the next subsection. The network uses 2D convolution layers, max-pooling layers, and fully-connected layers with rectified linear units (ReLU, \cite{Nair2010}) and sigmoid function as activation and batch normalizations to normalize output values from each layer. Finally, the network’s output values are transformed inversely using the maximum and minimum values used for standardization. Table \ref{tab:tab1} shows the detailed structure of the network. Generally, in deep CNNs, layers are arranged in the order of convolution, batch normalization, ReLU, and max-pooling. In this study, the order is convolution, max-pooling, batch normalization, and ReLU because accuracy in validation data is better than the standard order.

The network is trained separately for each forecast lead time using Adam \citep{Kingma2015} as the optimizer and mean square error as a loss function. The network is not stratified by season because the proposed method improves forecast accuracy without using seasonal stratification, as shown in section 4. The dataset period is from 00 UTC 8 October 2010 to 21 UTC 31 December 2019, at three-hourly intervals based on data stored at JMA. The dataset is divided into three parts: training period, from 8 October 2010 to 31 December 2017; validation period, from 1 January 2018 to 31 December 2018; and test period, from 1 January 2019 to 31 December 2019. In some previous studies, especially in image classifications, training and validation data share the same period, and data points are selected randomly from the period. However, in this study, data points are divided to prevent overfitting because meteorological data generally have a strong temporal correlation. In verification of statistical post-processing, cross-validations such as a leave-one-year-out or leave-one-month-out cross-validation are used especially when the length of the data period is insufficient. However, the study uses a considerably long period of data from 2010 to 2019, so that cross-validations are not used because they could conceal a long-term trend during the period.

The network is trained using the training dataset, where data points are selected randomly with a mini-batch size of 20 in each epoch. Hyperparameters such as the combination of input variables, the network structure, the number of epochs, and the mini-batch size are adjusted according to the validation dataset. The length of the training period is selected by comparing different training periods: 2010–2017, 2013–2017, and 2016–2017, considering the influence of MSM updates since 2010. The prediction accuracy in the validation period showed that the longer the training period, the better the accuracy (figure not shown). Despite MSM being updated several times since 2010, the result suggests that the characteristics of MSM have not changed so much. Thus, this study uses the longest period from 2010 to 2017 as the training period. After the network’s hyperparameters are adjusted and the network is trained, the network’s prediction accuracy is verified using the test dataset independent of the training and validation dataset.

\subsection{Standardization}
\label{sec:sec3-3}

Appropriate standardization is a key to training a network efficiently (Sola and Sevilla 1997). If the value ranges of the input variables have a large gap between channels, the training process would be ineffective, resulting in poor forecast accuracy. A simple standardization, in which 256 gradations are transformed into a 0–1 value range, is common for a typical image processing task based on color or black-and-white images. However, the same standardization could not yield good results in this study because meteorological data do not have a 256-gradation range, e.g., the surface temperature ranges from approximately $-20~^\circ$C to $40~^\circ$C, and the MSLP ranges from approximately 980–1030 hPa in Japan. Maximum and minimum value standardization is a simple method to standardize meteorological data in a 0–1 value range, as shown in the equation below:

\begin{equation}
\phi' = \frac{\phi - \phi_{\mathrm{min}}}{\phi_{\mathrm{max}} - \phi_{\mathrm{min}}}
\end{equation}

where $\phi$ represents an original variable, $\phi'$ represents a standardized variable, and $\phi_{\mathrm{max}}$ and $\phi_{\mathrm{min}}$ represent the maximum and minimum of  during the training period, respectively. Some JMA operational guidance, such as precipitation amount, minimum humidity, and sunshine duration, have used the above method. The method may yield good results; however, there is room for improvement because the standardized variables do not distribute evenly between 0 and 1 but localize when the method is applied to data at a specific time.

In this study, the maximum and minimum value of NWP outputs at each targeted time are used for standardization, e.g., MSLP at a targeted time is standardized using the maximum and minimum MSLP in $64\times 64$ grid points at the time, so the standardized variables are distributed evenly between 0 and 1. However, temperatures are standardized, with $+3$ K as the maximum and $-3$ K as the minimum, to consider extreme hot or cold temperatures, which were unpredictable using NWP models.

As shown in Fig. \ref{fig:fig2}, the maximum and minimum values of the NWP surface temperatures used for standardization are applied again when output values are transformed inversely. The estimated temperature used as the network’s ground truth is also standardized on the basis of the maximum $+3$ K and minimum $-3$ K value of input NWP surface temperatures to calculate the loss function.

\section{Verification results}
\label{sec:sec4}

\begin{figure}[b]
  \centering
  \includegraphics[width=0.9\textwidth]{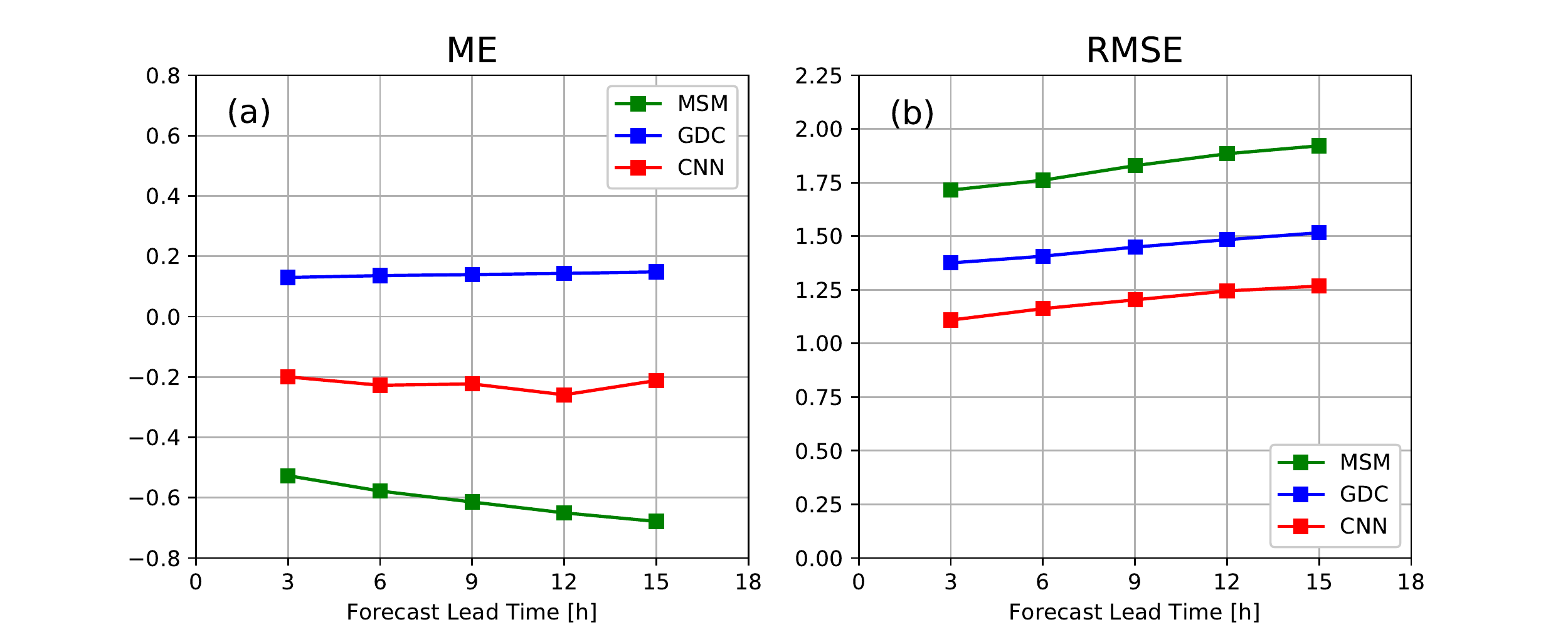}
  \caption{Verification results of MSM, GDC and CNN temperature forecast. (a) Mean error and (b) root mean square error for forecast times from 3 to 15 h. The test period is from 1 January 2019 to 31 December 2019.}
  \label{fig:fig3}
\end{figure}

This section presents the verification results of the network’s temperature prediction. In comparison, predictions of MSM surface temperature, an input variable to the network, and operational gridded temperature guidance are shown. The temperature of the estimated weather distribution is used as the ground truth of the verification. The verification metrics are the mean error (ME) and root mean square error (RMSE) defined as follows:

\begin{eqnarray*}
\mathrm{ME} = \frac{1}{T} \sum^{T}_{t=1}\frac{1}{N}\sum^{N}_{n=1}\left( F_{nt} - O_{nt} \right) \\
\mathrm{RMSE} = \sqrt{ \frac{1}{T} \sum^{T}_{t=1}\frac{1}{N}\sum^{N}_{n=1}\left( F_{nt} - O_{nt} \right)^2 }
\end{eqnarray*}

where $T$ and $N$ are the numbers of times and grid points used for the verification, respectively, and $F_{nt}$ and $O_{nt}$ are predicted and estimated temperature at a point $n$ and time $t$, respectively.

Figure \ref{fig:fig3} shows ME and RMSE for forecast lead times from 3 to 15 h. The test period is 1 year, from 1 January 2019 to 31 December 2019. The green, blue, and red lines and dots represent verification results of MSM, the operational gridded temperature guidance (GDC), and the prediction by the network (CNN), respectively. Since the estimated weather distribution has been defined only on land, the metrics are calculated on land also. As shown in Fig. \ref{fig:fig3}, the CNN model greatly reduces the RMSE from MSM and GDC. Hence, the network greatly improves operational gridded temperature guidance and could improve operational snowfall amount and precipitation type guidance. The ME shows that the CNN model has a negative bias of approximately $-0.2~^\circ$C, caused by a long-term trend of increasing negative bias of MSM surface temperature during the dataset period used in this study (figure not shown). The network cannot correct the long-term trend because it does not input any variables related to the date.

\section{Case studies}
\label{sec:sec5}

In the following subsections, three notable cases that occurred during the test period are shown. The first case is a low-temperature case associated with radiative cooling. It typically happens on clear and calm nights in winter. It is difficult to predict a sudden temperature drop by radiative cooling because the operational temperature guidance based on the Kalman filter cannot follow a rapid change of meteorological conditions. The second demonstrates a case with a positional error of a coastal front, which sometimes happens in plains in Japan, facing the Pacific Ocean. The positional error can lead to a large temperature difference between the forecast and observation. Lastly, an intense heat case in which MSM has a large negative temperature bias in summer due to the excessive upper-level cloud coverage \citep{Kusabiraki2020} is shown.

\subsection{Low-temperature case associated with radiative cooling}
\label{sec:sec5-1}

\begin{figure}[t]
  \centering
  \includegraphics[width=0.68\textwidth]{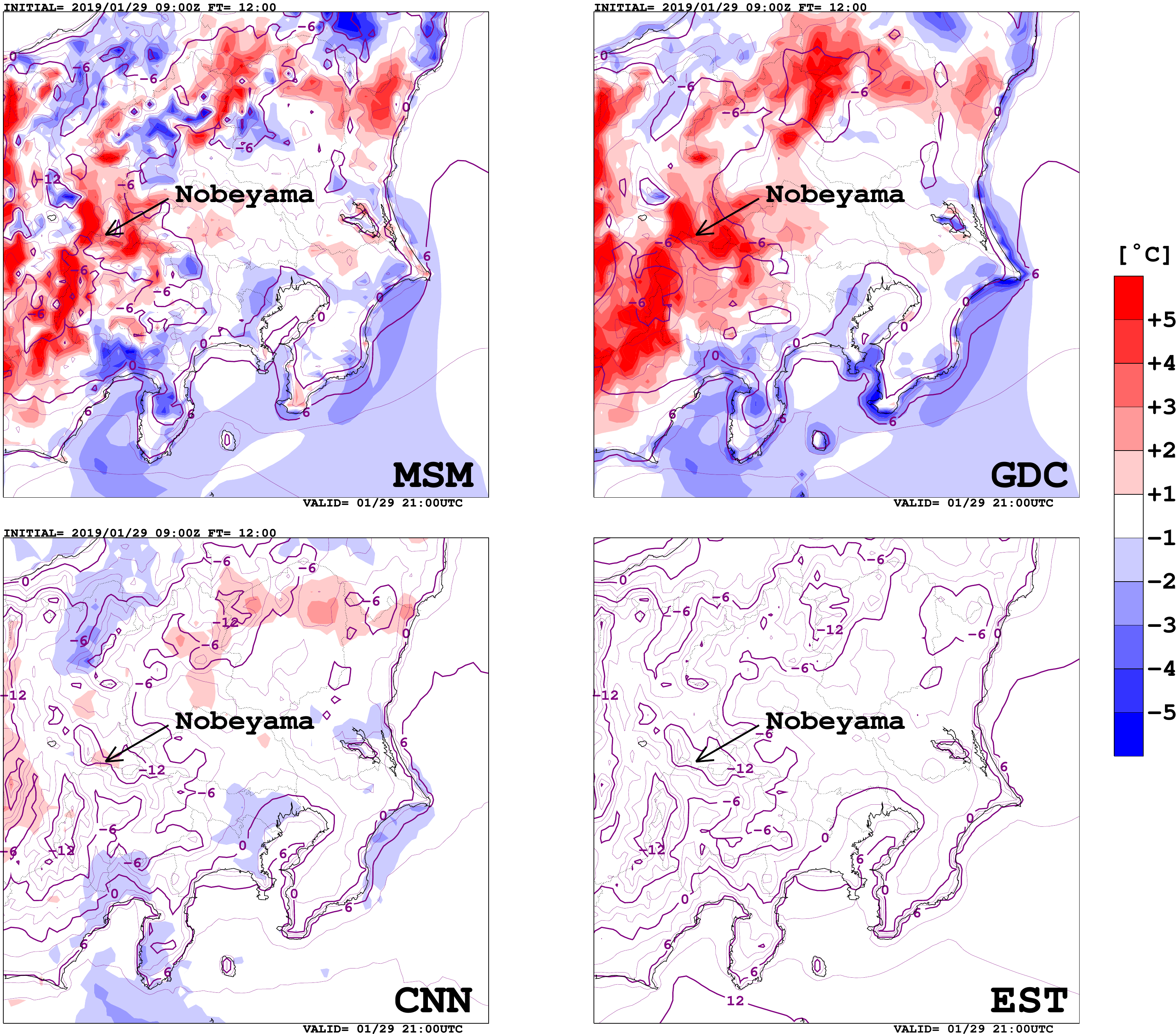}
  \caption{Temperature forecast (contours) at 21 UTC on 29 January 2019, by MSM (upper left), the gridded temperature guidance (GDC, upper right), and the CNN model (lower left) initialized at 09 UTC on 29 January 2019. The lower right is the estimated weather distribution (EST) surface temperature at 21 UTC on 29 January 2019. The color shades indicate the difference between the predicted and estimated temperatures.}
  \label{fig:fig4}
\end{figure}

\begin{figure}[t]
  \centering
  \includegraphics[width=0.5\textwidth]{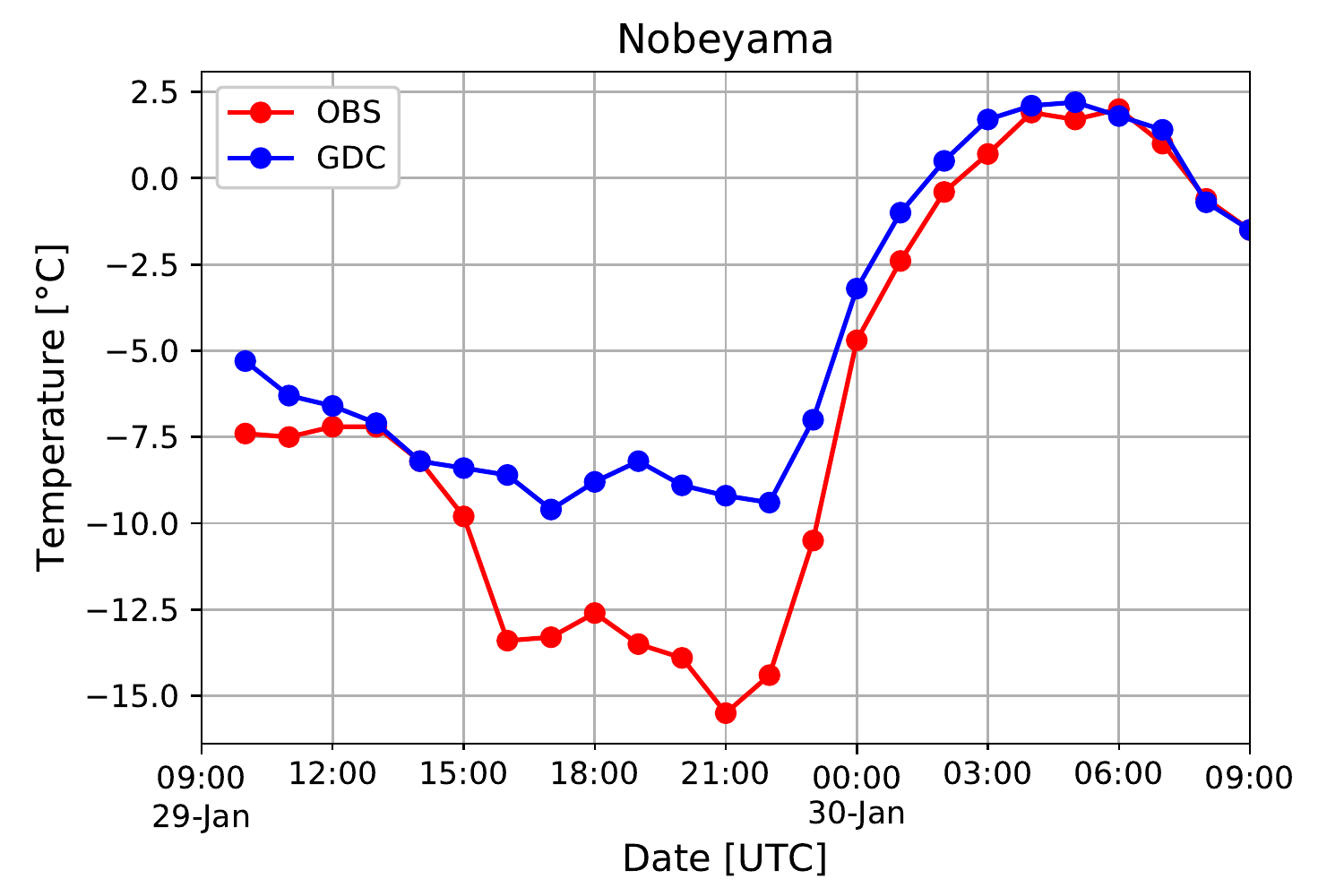}
  \caption{Operational point-like temperature guidance (GDC) at Nobeyama initialized at 09 UTC on 29 January 2019, and temperature observation (OBS) from 09 UTC 29 January 2019 to 09 UTC 30 January 2019.}
  \label{fig:fig5}
\end{figure}

\begin{figure}[t]
  \centering
  \includegraphics[width=0.88\textwidth]{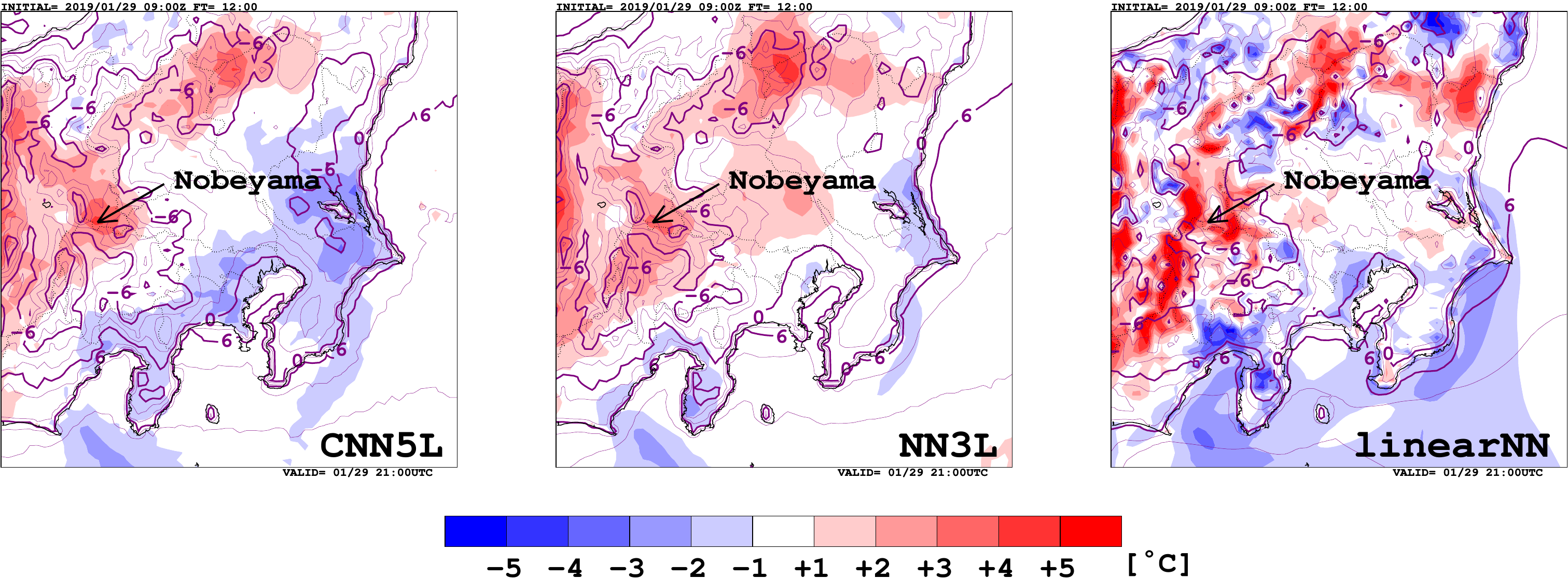}
  \caption{The same as Fig. \ref{fig:fig4} but the results of CNN5L, NN3L, and linearNN from the left to the right panel, respectively.}
  \label{fig:fig6}
\end{figure}

Early morning on 30 January 2019 (LT), a high-pressure system covered the Kanto region (figure not shown). It was almost clear and calm, which is prone to low temperatures associated with radiative cooling. Figure \ref{fig:fig4} shows predicted and estimated temperatures targeted at 21 UTC on 29 January 2019. The estimated weather distribution (EST) shows a low temperature, i.e., $-12^\circ$C or less, in the mountainous region (Fig. \ref{fig:fig1}). The other panels in Fig. \ref{fig:fig4} indicate the 12-h forecast temperature by MSM, the gridded temperature guidance (GDC), and the CNN model initialized at 09 UTC on 29 January 2019, along with the difference between the predicted and estimated temperatures. The red and blue shades indicate that the predicted temperature was higher and lower than the estimation, respectively. The temperature predictions obtained by MSM and GDC were higher than EST in the mountainous region.

Figure \ref{fig:fig5} shows a time series of the observed temperature (OBS) and the operational point-like temperature guidance (GDC) initialized at 09 UTC on 29 January 2019, at Nobeyama (shown in Fig. \ref{fig:fig4}). The observed temperature was approximately $-7.5^\circ$C until 13 UTC on 29 January 2019; then, it dropped to $-15.5^\circ$C at 21 UTC on 29 January 2019. The temperature guidance could not follow such a low temperature, and the error was larger than $6^\circ$C at 21 UTC, which is consistent with Fig. \ref{fig:fig4}. A large temperature difference between guidance and observation sometimes happens under radiative cooling conditions and seems to be one of the greatest challenges to accurate temperature prediction.

The 12-h forecast temperature by the CNN model in Fig. \ref{fig:fig4} shows almost the same temperature as EST at Nobeyama and in the entire forecast area. The results would be due to the CNN model’s ability to represent spatial structures and nonlinear relationships between predictors and predictand. To show the effect of deep CNN operations, three ablation studies are performed. The first method is the five-layer CNN that removes two middle layers from the proposed CNN model, i.e., removes Conv2-layer and ConvT1-layer in Table \ref{tab:tab1} (hereafter, called CNN5L). The second method is the three-layer NN that removes two more middle layers from CNN5L, i.e., removes Conv1-layer and ConvT2-layer in Table \ref{tab:tab1} (hereafter, called NN3L). The last method removes all middle layers from the proposed CNN model and simply connects the input and output layer at each grid point with a linear activation function (hereafter, called linearNN), which could approximate multiple linear regression at each grid point. All three methods used the same settings including training and validation period except for the structure of the network, the number of epochs that was decided by the loss function toward validation dataset, and the number of units input into FC1-layer and output from FC2-layer, which was decided according to the structure of the network. Results of the ablation studies (Fig. \ref{fig:fig6}) showed that linearNN predicted almost the same as MSM in Fig. \ref{fig:fig4}, whereas CNN5L and NN3L predicted better than MSM and GDC around Nobeyama and mountainous region. The results suggest that the formidable prediction by CNN in Fig. \ref{fig:fig4} came from a deep CNN architecture considering both complex nonlinearity and spatial structure. Being diagnosed as the surface temperature in 5-km grids, MSM surface temperature has a bias from the temperature at actual height. The difference becomes large around valleys and mountain tops, and when the temperature lapse rate at the lower troposphere is close to the dry adiabatic lapse rate, such as at a time of radiative cooling. Because the CNN model uses surface and lower troposphere temperature along with MSLP and wind components as predictors, it can predict low temperature caused by radiative cooling accurately. The CNN model is of great significance in low-temperature cases during winter.

\begin{figure}[t]
  \centering
  \includegraphics[width=0.9\textwidth]{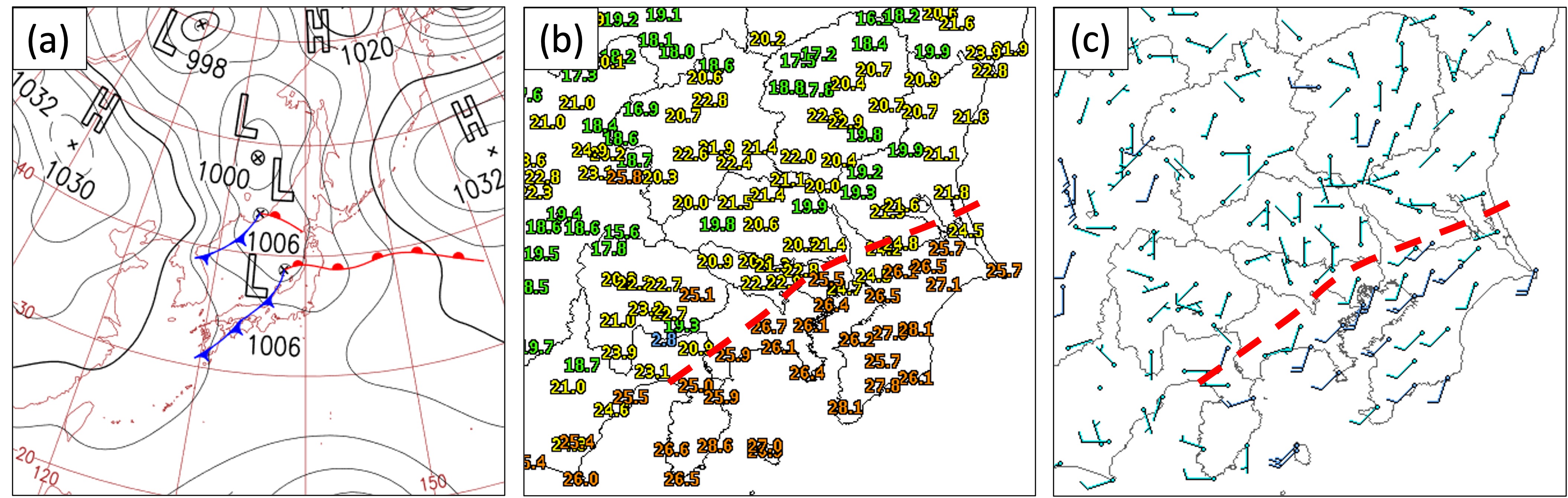}
  \caption{(a) Surface weather map at 00 UTC on 8 October 2019. (b, c) Surface temperature [$^\circ$C] and wind observations at AMeDaS sites at 03 UTC on 8 October 2019. The red dashed lines in (b) and (c) indicate the position of the coastal front.}
  \label{fig:fig7}
\end{figure}

\begin{figure}[t]
  \centering
  \includegraphics[width=0.68\textwidth]{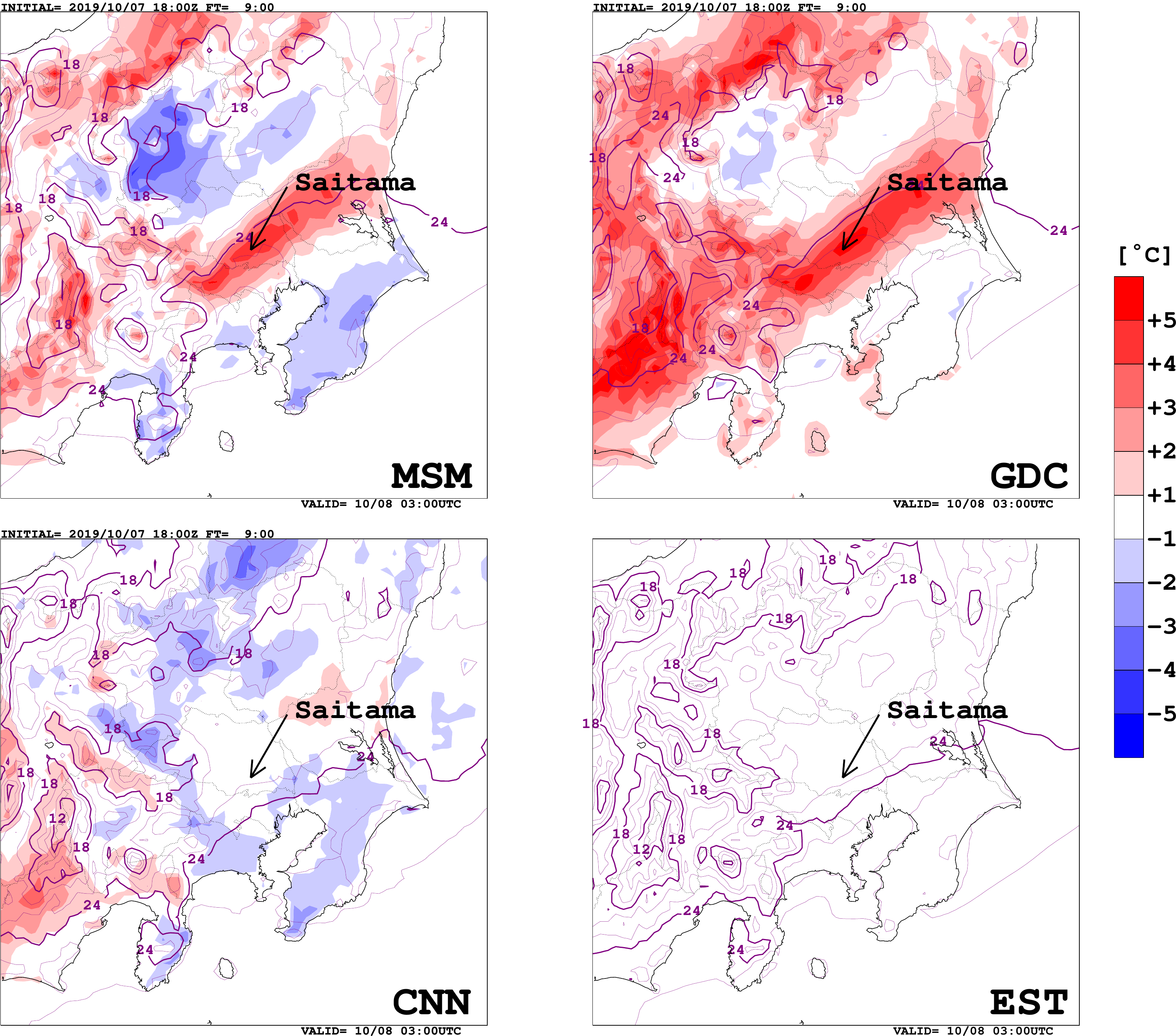}
  \caption{The same as Fig. \ref{fig:fig4} but the initial time at 18 UTC on 7 October 2019 and projection time at 03 UTC on 8 October 2019.}
  \label{fig:fig8}
\end{figure}

\subsection{Positional error of coastal front}
\label{sec:sec5-2}

\begin{figure}[t]
  \centering
  \includegraphics[width=0.5\textwidth]{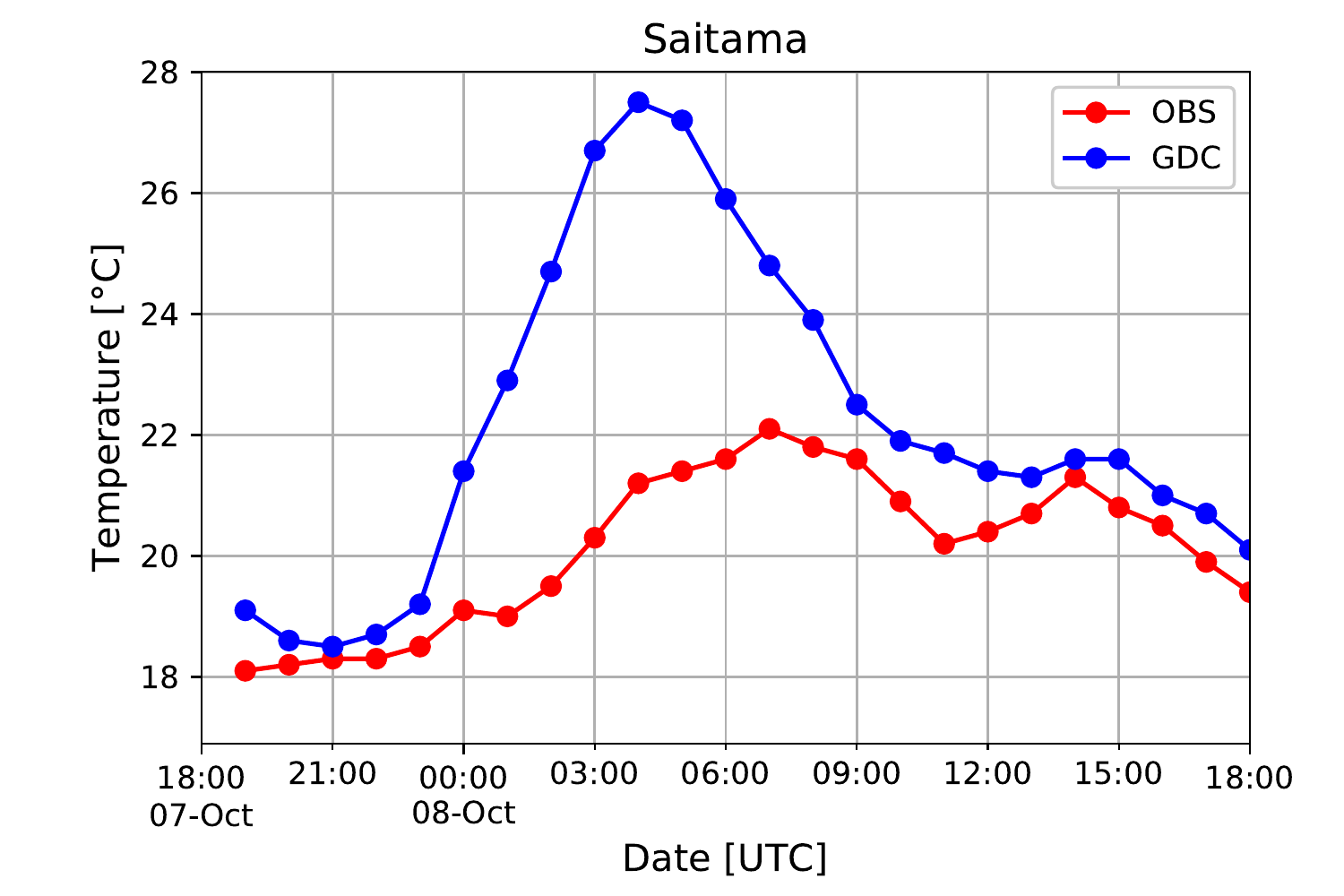}
  \caption{The same as Fig. \ref{fig:fig5} but at Saitama initialized at 18 UTC on 7 October 2019 and observation time from 18 UTC 7 October 2019 to 18 UTC 8 October 2019.}
  \label{fig:fig9}
\end{figure}

\begin{figure}[t]
  \centering
  \includegraphics[width=0.88\textwidth]{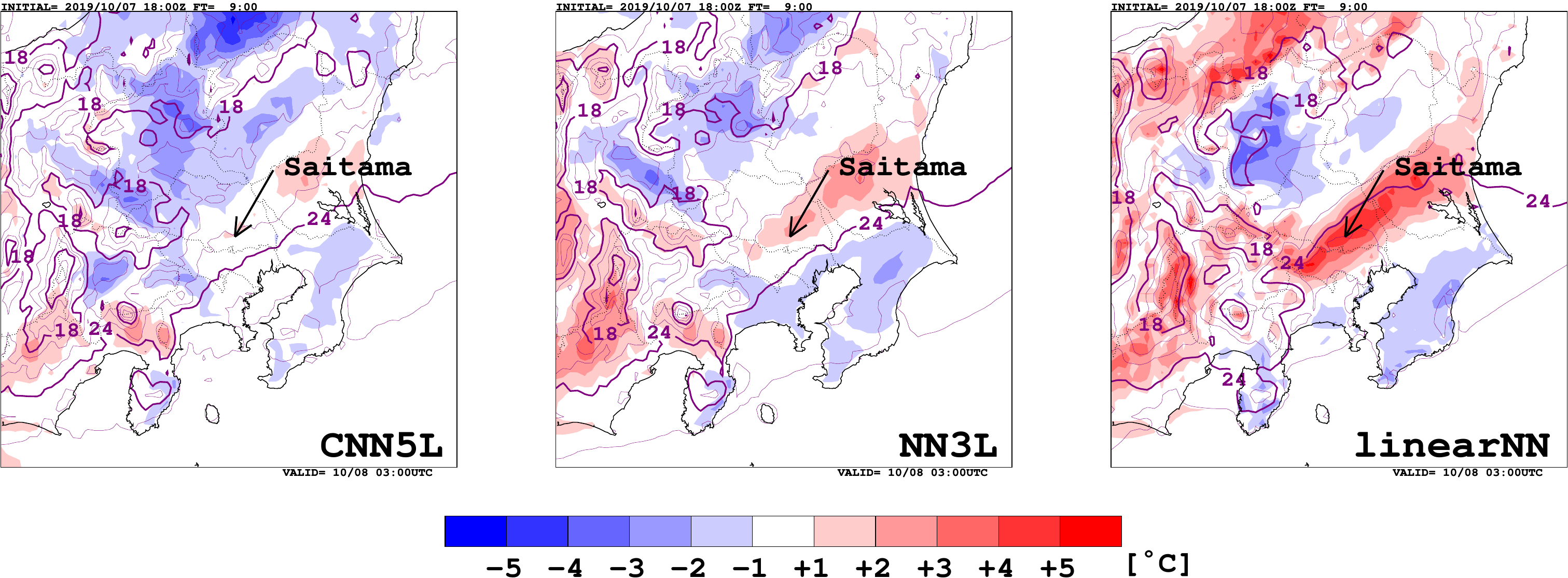}
  \caption{The same as Fig. \ref{fig:fig6} but the results of CNN5L, NN3L, and linearNN from the left to the right panel, respectively.}
  \label{fig:fig10}
\end{figure}

On 8 October 2019, a low-pressure system was located in the middle of the Sea of Japan (Fig. \ref{fig:fig7}a), and a warm south-westerly wind blew across the southern part of the Kanto region (Figs. \ref{fig:fig7}b and \ref{fig:fig7}c), which generated a coastal front indicated by the red dashed lines in the Figs. \ref{fig:fig7}b and \ref{fig:fig7}c. Temperatures in the south of the coastal front exceeded $25^\circ$C, whereas temperatures in the north hovered around $20^\circ$C. A large difference in temperature and wind direction caused by coastal fronts typically occurs in the Kanto region when there is a low-pressure system in the middle of the Sea of Japan.

Figure \ref{fig:fig8} shows 9-h forecasts and estimated temperature targeted at 03 UTC on 8 October 2019. The estimated $24^\circ$C isothermal line located in the south of the Kanto region corresponds to the coastal front in Fig. \ref{fig:fig7}. In this case, MSM predicted the coastal front approximately 30 km further north than the actual position, which results in the largest temperature difference around the coastal front. Preceding studies \citep{Hara2014, Kawano2019} have shown that MSM has a systematic error in predicting coastal fronts further north than the actual position. \cite{Suzuki2021} concluded that the positional error comes from the topography difference between reality and NWP models by performing sensitivity experiments using JMA Non-Hydrostatic Model (JMA-NHM, \cite{Saito2007}). Their study suggests that the positional error is a bias that statistical methods can remove. However, the operational guidance cannot remove the bias, and it is known as one of the most difficult phenomena to predict temperatures accurately \citep{Sannohe2018}.

Figure \ref{fig:fig9} shows a time series of the observed temperature and the operational point-like temperature guidance initialized at 18 UTC on 7 October 2019, at Saitama (shown in Fig. \ref{fig:fig8}). The observed temperature was below $22.2^\circ$C during the daytime, whereas the guidance predicted a temperature higher than $27^\circ$C around 04 UTC. The coastal front was on the south of Saitama throughout the day, whereas MSM predicted the position of the coastal front on the north. The temperature difference between the operational guidance and observation was greater than $6^\circ$C at 03 UTC.

The 9-h forecast by the CNN model (Fig. \ref{fig:fig8}) predicted the position of the coastal front, or a $24^\circ$C isothermal line, successfully, so the temperature was almost the same as EST around Saitama. Results of ablation studies (Fig. \ref{fig:fig10}) showed that the deeper the network, the better the accuracy around the coastal front, which suggests considering both nonlinearity and spatial structure via deep CNN brings about the accurate prediction in this case. As shown in \cite{Suzuki2021}, insufficient topography in MSM brings about the positional error of coastal fronts. CNN can compensate for the deficiency by using temperature information affected by topography around the coastal front through convolution, pooling, and fully-connected layers.

\subsection{Intense heat case}

\begin{figure}[t]
  \centering
  \includegraphics[width=0.68\textwidth]{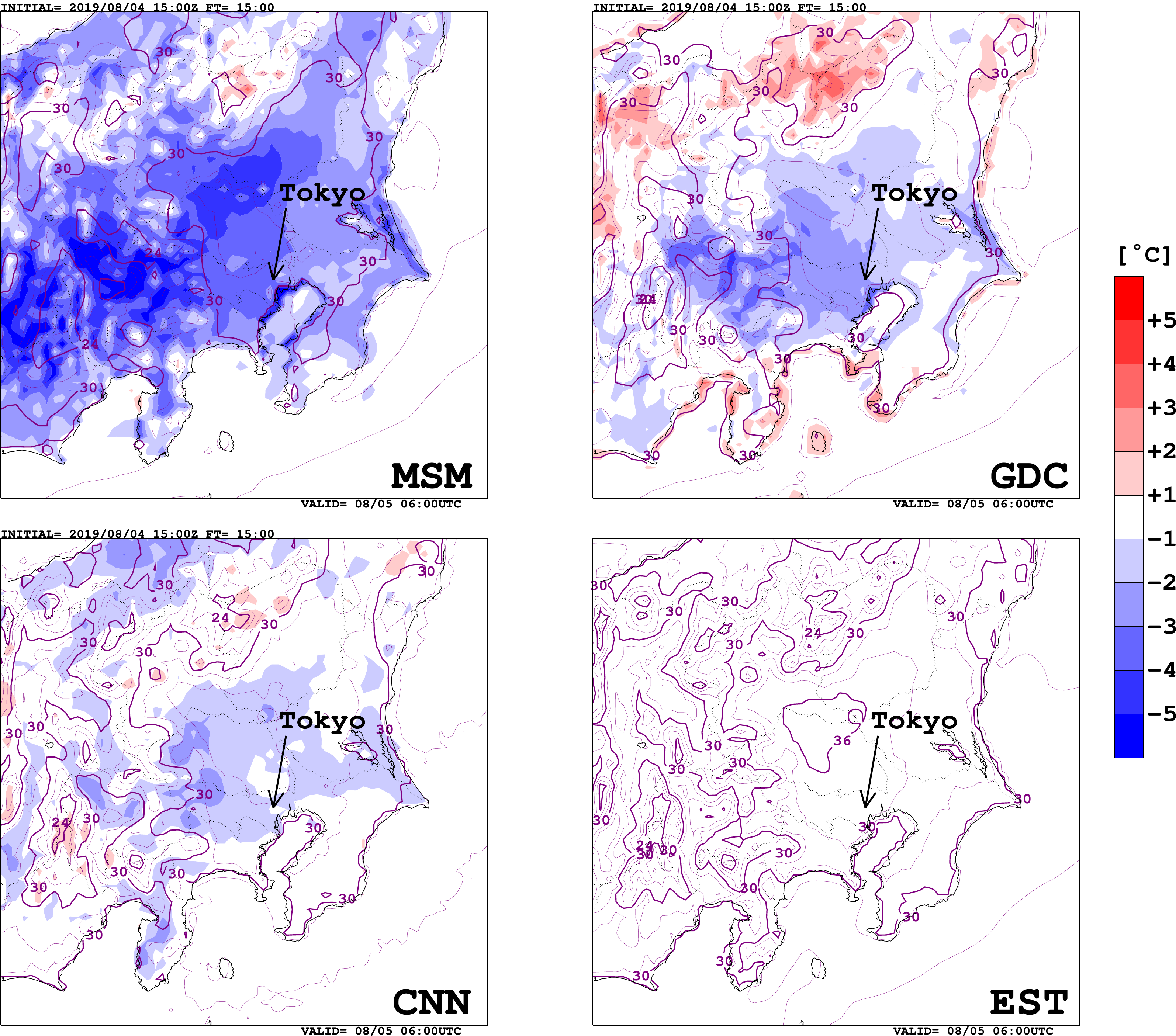}
  \caption{The same as Fig. \ref{fig:fig4} but the initial time at 15 UTC on 4 August 2019 and projection time at 06 UTC on 5 August 2019.}
  \label{fig:fig11}
\end{figure}

\begin{figure}[t]
  \centering
  \includegraphics[width=0.5\textwidth,clip]{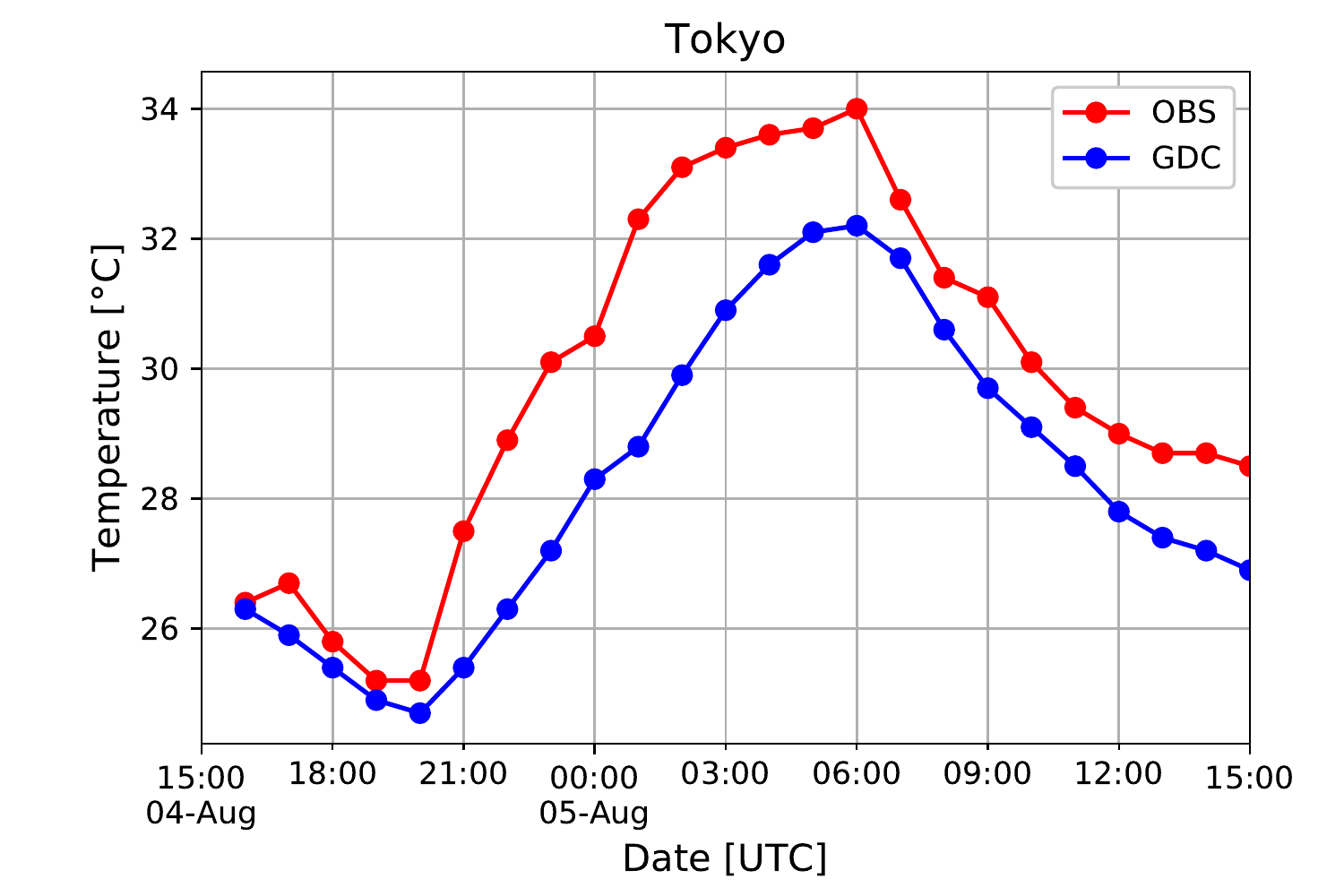}
  \caption{The same as Fig. \ref{fig:fig5} but at Tokyo initialized at 15 UTC on 4 August 2019 and observation time from 15 UTC 4 August 2019 to 15 UTC 5 August 2019.}
  \label{fig:fig12}
\end{figure}

\begin{figure}[t]
  \centering
  \includegraphics[width=0.88\textwidth]{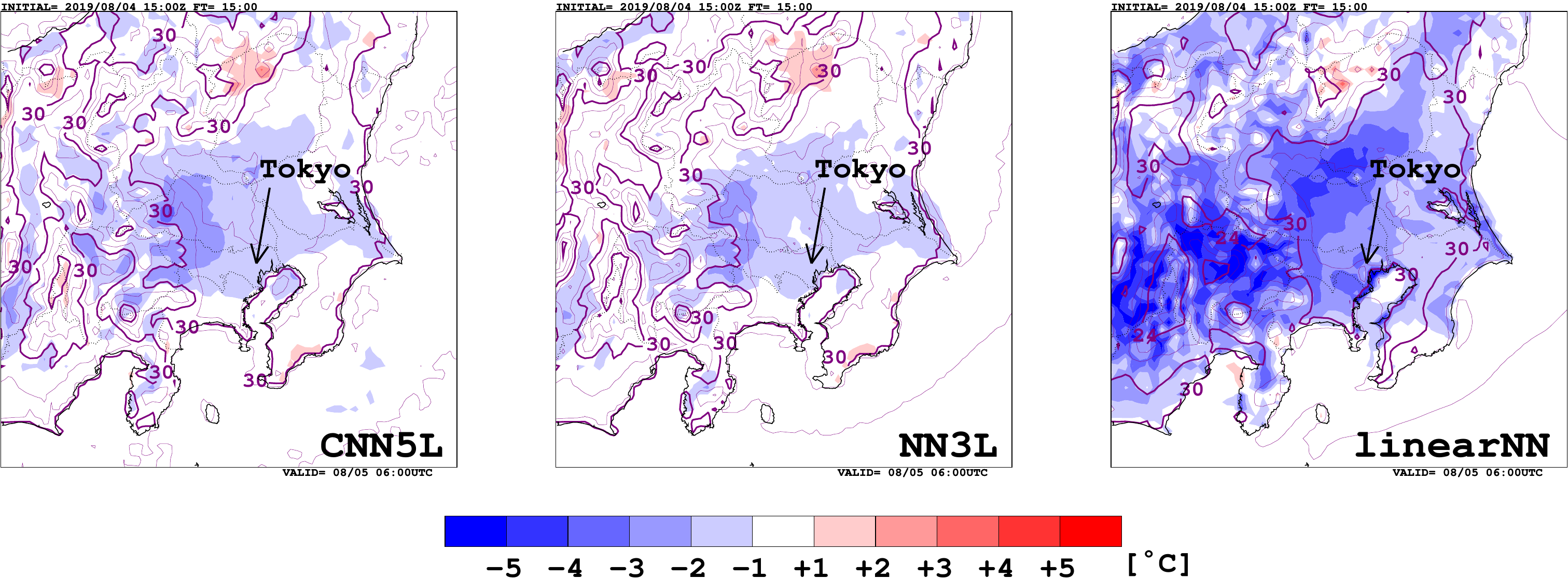}
  \caption{The same as Fig. \ref{fig:fig11} but the results of CNN5L, NN3L, and linearNN from the left to the right panel, respectively.}
  \label{fig:fig13}
\end{figure}

\label{sec:sec5-3}
The maximum temperatures at the inland area of the Kanto region were higher than $35^\circ$C on 5 August 2019 because the Pacific anticyclone had covered Japan for several days, and a typhoon located in the south of Japan brought warm southerly wind (figure not shown). Figure \ref{fig:fig11} shows 15-h forecasts and estimated temperature targeted at 06 UTC on 5 August 2019. MSM predicted temperatures lower than EST in almost all areas shown in the figure. GDC corrected MSM temperature to a certain degree, but it was lower than $2^\circ$C or more around Tokyo.

Figure \ref{fig:fig12} shows a time series of the observed temperature and operational point-like temperature guidance initialized at 15 UTC on 4 August 2019, at Tokyo (shown in Fig. \ref{fig:fig11}). The observed temperature was approximately $2^\circ$C higher than the operational guidance in the morning to the afternoon (from 21 UTC 4 August to 06 UTC 5 August).

The 15-h forecast by the CNN model (Fig. \ref{fig:fig11}) predicted high temperatures better than GDC, and the difference in temperatures between CNN and EST at 06 UTC was approximately $1^\circ$C around Tokyo. Results of ablation studies (Fig. \ref{fig:fig13}) showed that CNN5L and NN3L predicted similar to the CNN model, whereas linearNN predicted almost the same as MSM in Fig. \ref{fig:fig11}, which suggests CNN’s accurate prediction came from NN’s ability to represent nonlinear relationships between input and output variables. As shown in this case, MSM has a large negative bias for daytime temperatures in summer due to the excessive upper-level cloud coverage and subsequent insufficient downward shortwave radiation at the surface \citep{Kusabiraki2020}. CNN can effectively correct the negative bias by using temperature information in the target area because it might be a simple bias that frequently occurs in summer.

\section{Conclusions}
\label{sec:sec6}

This study demonstrated gridded temperature predictions on the basis of an encoder–decoder-based CNN around the Kanto region. The network has seven layers with 2D convolution layers. Seven NWP output variables, including temperature at four height levels, wind components, and MSLP, are used as input, and gridded temperatures by the estimated weather distribution are used as the ground truth of the network. Input variables are standardized using the maximum and minimum values at each targeted time, ensuring that the variables are evenly distributed between 0 and 1, which is crucial for effectively training the network.

Verification results using an independent dataset from training and validation datasets showed that the CNN model greatly improved operational gridded temperature guidance. Additionally, the CNN model can predict temperatures associated with radiative cooling, coastal fronts, and heatwaves, which have been difficult to correct using operational temperature guidance. Operational temperature forecast accuracy will be greatly improved using the network demonstrated in this study. The operational snowfall amount and precipitation type guidance will also be improved because these elements use gridded temperature guidance as input variables.

Future issues of the study are as follows: to extend the method to other areas in Japan; to extend the forecast range after 15 h; to improve prediction accuracy using other predictors and/or networks; to find a suitable stratification; to correct the long-term trend using variables related to a specific date; and to apply the network to other elements such as wind, weather category, precipitation amount, and probability forecasts.

\bibliographystyle{unsrtnat}


\end{document}